\documentclass{jfm}
\usepackage{graphicx}
\usepackage{amsmath,amssymb}
\usepackage{epstopdf, epsfig}

\shorttitle{Two-point correlation in wall turbulence}
\shortauthor{H. Mouri}

\title{Two-point correlation in wall turbulence according to the attached-eddy hypothesis}

\author{Hideaki Mouri
       }

\affiliation{Meteorological Research Institute, Nagamine, Tsukuba 305-0052, Japan}

\begin{document}

\maketitle

\begin{abstract}
For the constant-stress layer of wall turbulence, two-point correlations of velocity fluctuations are studied theoretically by using the attached-eddy hypothesis, i.e., a phenomenological model of a random superposition of energy-containing eddies that are attached to the wall. While the previous studies had invoked additional assumptions, we focus on the minimum assumptions of the hypothesis to derive its most general forms of the correlation functions. They would allow us to use or assess the hypothesis without any effect of those additional assumptions. We also study the energy spectra and the two-point correlations of the rate of momentum transfer and of the rate of energy dissipation.
\end{abstract}

\section{Introduction} \label{sec1}
\begin{figure}
  \centerline{\resizebox{6cm}{!}{\includegraphics*[2.5cm,13.5cm][18.5cm,23.5cm]{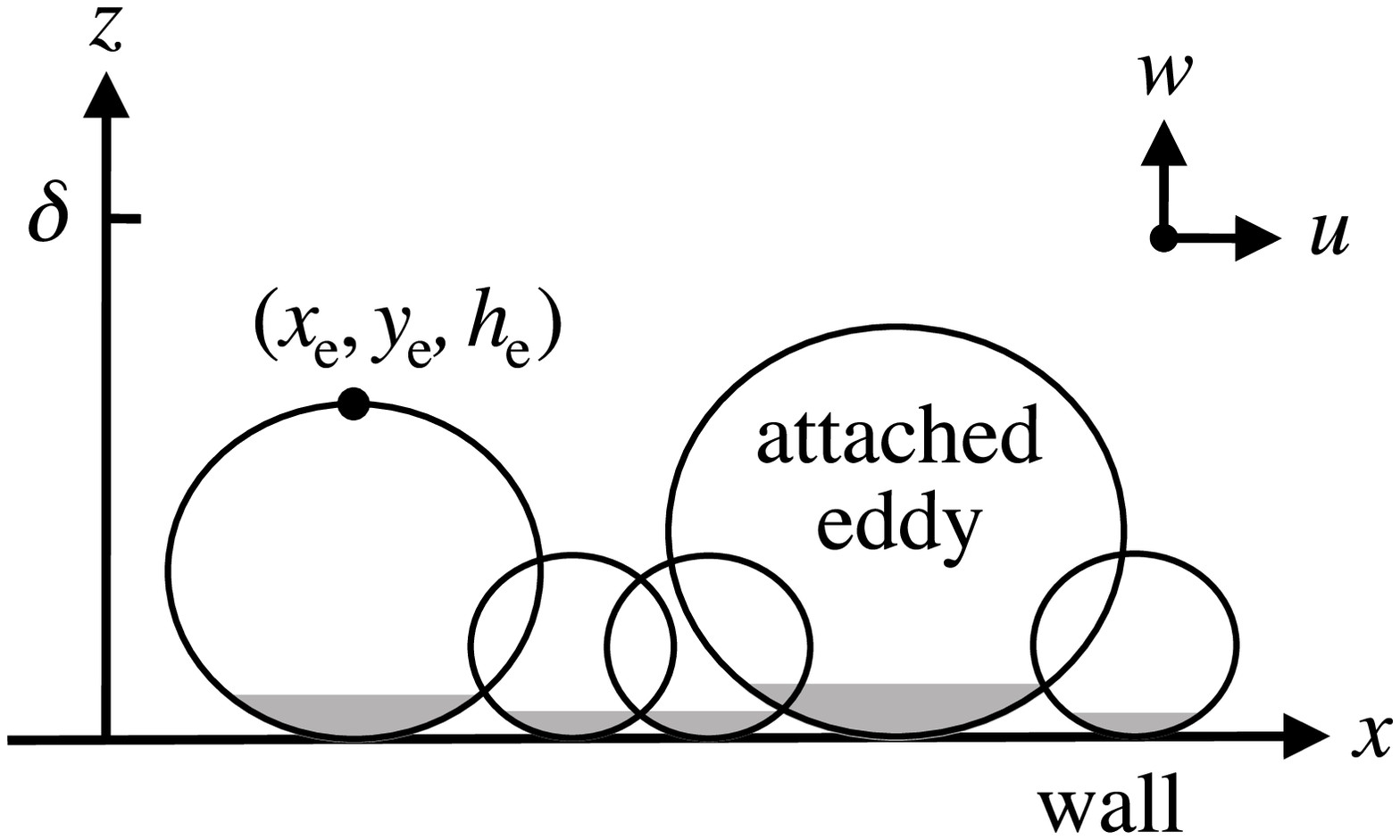}}}
  \caption{Schematic of attached eddies in wall turbulence, where grey areas denote the undermost layers of the eddies}
\label{fig1}
\end{figure}

This is a theoretical study about wall turbulence that is formed in a pipe, in a channel, or over a plate. As shown in figure \ref{fig1}, we set a smooth wall at the $x$-$y$ plane, set the mean stream along the $x$ direction, and use $u(z)$, $v(z)$, and $w(z)$ to denote velocity fluctuations in the streamwise, spanwise, and wall-normal directions at a height $z$ from the wall. The mean streamwise velocity itself is not studied here. If the turbulence is stationary and at a high Reynolds number, it has a layer with some constant value of $\rho u_{\tau}^2$ for the mean rate of momentum transfer, i.e., for the Reynolds stress $\rho \langle -uw \rangle$. Here $\rho$ is the mass density, $u_{\tau}$ is the friction velocity, and $\langle \cdot \rangle$ denotes the ensemble average.

For this constant-stress layer, there is a phenomenological model of a random superposition of energy-containing eddies that are attached to the wall, i.e., the attached-eddy hypothesis (\cite{t76}). The velocity fields of the eddies are set to have an identical shape with a common characteristic velocity $u_{\tau}$, while their sizes are distributed without any characteristic size. Actually, the constant-stress layer has a characteristic constant $u_{\tau}$ in units of velocity but has no characteristic constant in units of length. If the size distribution of the attached eddies is given so as to reproduce the constantness of $\langle -uw \rangle$, a logarithmic law is predicted for the variance of the streamwise velocity fluctuations $u(z)$ as
%____________________________________________________
\begin{equation}
\label{eq1_1}
\frac{\langle u^2(z_1) \rangle -\langle u^2(z_2)\rangle}{u_{\tau}^2} \varpropto \ln \! \left( \frac{z_1}{z_2} \right).
\end{equation}
%____________________________________________________
Since this prediction has been confirmed in experiments of a variety of wall turbulence (\cite{hvbs12,marusic13}), the attached-eddy hypothesis has turned out to be a reliable model.

The important application of such a hypothesis would be to the two-point correlations or equivalently to the energy spectra of the velocity fluctuations. Besides numerical works based on a particular model of the eddies (e.g., \cite{marusic01}), there are some theoretical studies (e.g., \cite{pc82,dnk06}). They have invoked, however, additional assumptions such as for the similarity (\cite{pa77}). We rather focus on the minimum assumptions of the attached-eddy hypothesis that have led to the law of equation (\ref{eq1_1}). These assumptions are yet sufficient to constrain the correlation functions and to clarify their most general forms, which would offer an opportunity to use or assess the hypothesis without any effect of the additional assumptions. We also study the energy spectra and the two-point correlations of some other quantities. They are to be compared with the existing models of the wall turbulence.

%____________________________________________________
%____________________________________________________

\section{Basic setting of the hypothesis} \label{sec2}

Here is a summary of the attached-eddy hypothesis (\cite{t76}). The turbulence is set homogeneous in the streamwise and spanwise directions. For the case of a boundary layer, we assume that it has been well developed and hence it is negligibly dependent on the streamwise position $x$. Since this hypothesis is for the constant-stress layer, the value of the kinematic viscosity $\nu$ is not essential. The limit $\nu \rightarrow 0$ is taken so as to ignore the viscous length $\nu/u_{\tau}$ with respect to the height $z$. Then, a free-slip condition, i.e., $u \ne 0$ and $v \ne 0$, is imposed on the wall at $z = 0$.

The attached eddies are extending from the wall into the flow but are bounded in each of the directions. Regardless of the size of the eddies, they have an identical shape with a common characteristic velocity $u_{\tau}$. That is, if $\mbox{\boldmath{$x$}}_{\rm e} = (x_{\rm e}, y_{\rm e}, h_{\rm e})$ lies at the highest position of an eddy (see figure \ref{fig1}), its velocity field $\mbox{\boldmath{$v$}}_{\rm e}$ is given for a position $\mbox{\boldmath{$x$}} = (x, y, z)$ as
%____________________________________________________
\begin{subequations}
\begin{equation}
\label{eq2_1a}
\frac{\mbox{\boldmath{$v$}}_{\rm e}(\mbox{\boldmath{$x$}})}{u_{\tau}}
=
\mbox{\boldmath{$f$}} \! \left( \frac{\mbox{\boldmath{$x$}} - \mbox{\boldmath{$x$}}_{\rm e} }{h_{\rm e}} \right)
\quad \mbox{with} \ \
\mbox{\boldmath{$f$}} = (f_u, f_v, f_w). 
\end{equation}
%____________________________________________________
We regard $h_{\rm e}$ as the size of this eddy. The existence of the wall imposes $f_w = 0$ at $z = 0$. As for the undermost layer at $z \ll h_{\rm e}$ of such an eddy, it is enough to assume $f_w \varpropto z/h_{\rm e}$. Furthermore, the free-slip wall condition imposes $f_u \ne 0$ and $f_v \ne 0$ at $z = 0$. These are summarized by using $\mbox{\boldmath{$g$}}$ as some function of $(x-x_{\rm e})/h_{\rm e}$ and $(y-y_{\rm e})/h_{\rm e}$,
%____________________________________________________
\begin{equation}
\label{eq2_1b}
f_u \rightarrow                  g_u, \ \
f_v \rightarrow                  g_v, \ \   \mbox{and} \ \ 
f_w \rightarrow \frac{z}{h_e} \, g_w  \quad \mbox{as}  \ \ \frac{z}{h_e} \rightarrow 0.
\end{equation}
\end{subequations}
%____________________________________________________
The condition at $z > h_e$ is $f_u = f_v = f_w = 0$. No more assumption is required about the functional form of $\mbox{\boldmath{$f$}}$.

The sizes of the attached eddies are distributed continuously from $h_{\rm e} = \nu /u_{\tau} \rightarrow 0$ to $h_{\rm e} = \delta$. Here $\delta$ corresponds to the height of the wall turbulence, i.e., pipe radius, channel half-width, or boundary layer thickness. The asymptotic laws for $z/\delta \rightarrow 0$ are regarded as those for the constant-stress layer. On the wall, the eddies are distributed randomly and independently. They could even overlap one another. We do not assume any more about the distribution of the eddies, although some previous studies had assumed a particular hierarchy in that distribution (e.g., \cite{pc82}).

From the random and independent distribution of the attached eddies, it follows that the entire velocity field is a superposition of those of the individual eddies. Any two-point correlation over a streamwise distance $\mbox{\boldmath{$r$}}_l = \mbox{\boldmath{$r$}}_x = (r,0,0)$ at a height $z$ is the sum of those within the individual eddies of various sizes from $h_{\rm e} = z$ to $h_{\rm e} = \delta$,
%____________________________________________________
\begin{equation}
\frac{\langle v_i(\mbox{\boldmath $x$}+\mbox{\boldmath $r$}_x) v_j(\mbox{\boldmath $x$}) \rangle}{u_{\tau}^2} \nonumber \\ 
= \!
\int^{\delta}_z \! \frac{dh_{\rm e}}{h_{\rm e}}
\left[
        h_{\rm e}^3 \, n_{\rm e}(h_{\rm e})
        \! \int \! \frac{dx_{\rm e}}{h_{\rm e}} \! \int \! \frac{dy_{\rm e}}{h_{\rm e}} \, f_{v_i} \! \left( \frac{\mbox{\boldmath{$x$}} + \mbox{\boldmath{$r$}}_x - \mbox{\boldmath{$x$}}_{\rm e} }{h_{\rm e}} \right)
                                                                                           f_{v_j} \! \left( \frac{\mbox{\boldmath{$x$}}                           - \mbox{\boldmath{$x$}}_{\rm e} }{h_{\rm e}} \right) 
\right].
\end{equation}
%____________________________________________________
The subscripts $i$ and $j$ denote either of $x$, $y$, or $z$ such that $v_i$ and $v_j$ denote either of $u$, $v$, or $w$. As justified later, the number density of the attached eddies at size $h_{\rm e}$ per unit area of the wall is $n_{\rm e}(h_{\rm e}) = N_{\rm e} h_{\rm e}^{-3}$, where $N_{\rm e}$ is a constant. By using $\langle v_i(x+r,z)v_j(x,z) \rangle$ in place of $\langle v_i(\mbox{\boldmath $x$}+\mbox{\boldmath $r$}_x) v_j(\mbox{\boldmath $x$}) \rangle$,
%____________________________________________________
\begin{subequations}
\label{eq2_2}
\begin{equation}
\label{eq2_2a}
\frac{\langle v_i(x+r,z)v_j(x,z) \rangle}{u_{\tau}^2}
=
N_{\rm e} \! \int^{\delta}_z \! \frac{dh_{\rm e}}{h_{\rm e}} \, I_{v_i v_j} \! \left( \frac{r}{h_{\rm e}}, \frac{z}{h_{\rm e}} \right),
\end{equation}
%____________________________________________________
with the correlation for eddies of the size $h_{\rm e}$,
%____________________________________________________
\begin{equation}
\label{eq2_2b}
I_{v_i v_j} \! \left( \frac{r}{h_{\rm e}}, \frac{z}{h_{\rm e}} \right) 
= \!
\int \! \frac{dx_{\rm e}}{h_{\rm e}} \! \int \! \frac{dy_{\rm e}}{h_{\rm e}} \,
f_{v_i} \! \left( \frac{\mbox{\boldmath{$x$}} + \mbox{\boldmath{$r$}}_x - \mbox{\boldmath{$x$}}_{\rm e} }{h_{\rm e}} \right)
f_{v_j} \! \left( \frac{\mbox{\boldmath{$x$}}                           - \mbox{\boldmath{$x$}}_{\rm e} }{h_{\rm e}} \right) .
\end{equation}
\end{subequations}
%____________________________________________________
Since the streamwise and spanwise sizes of each eddy have been set finite, the integral is always finite. Together with the wall condition of equation (\ref{eq2_1b}), which yields a condition for $I_{v_i v_j}(r/h_{\rm e}, z/h_{\rm e})$ in case of $r/h_{\rm e} = 0$, equation (\ref{eq2_2}) serves as the basis of the attached-eddy hypothesis.

%____________________________________________________
\begin{figure}
  \centerline{\resizebox{13.8cm}{!}{\includegraphics*[0.9cm,20.0cm][19.2cm,26.0cm]{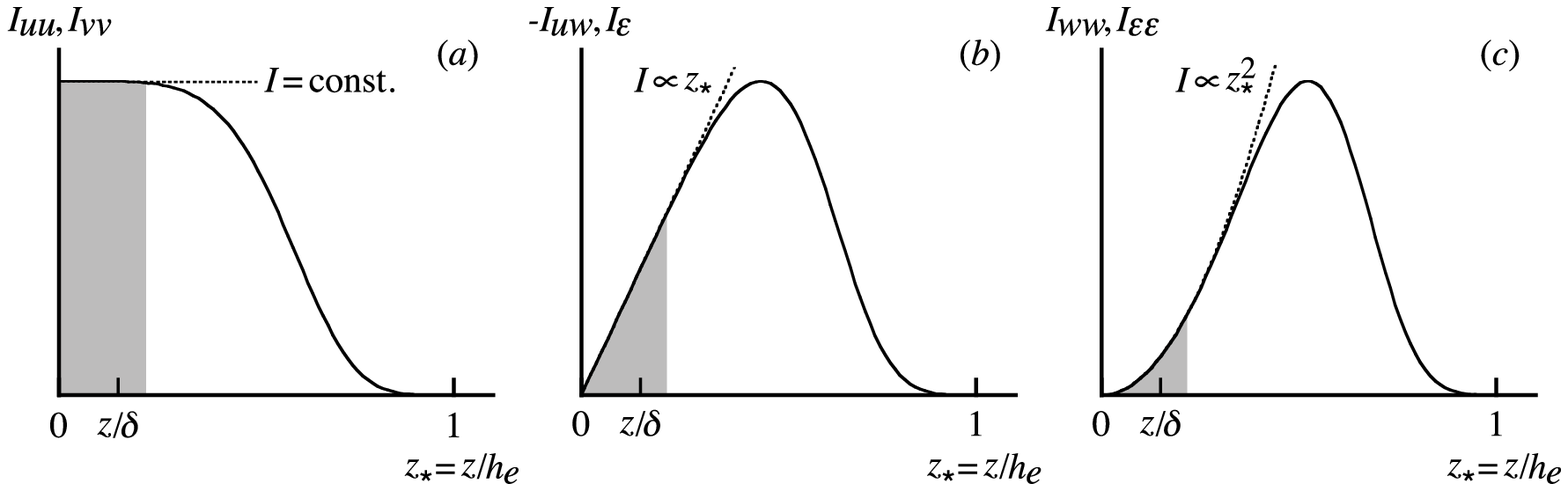}}}
  \caption{Schematic of $I_{uu}$, $I_{vv}$ ($a$), $I_{uw}$, $I_{\varepsilon}$ ($b$), and $I_{ww}$, $I_{\varepsilon \varepsilon}$ ($c$) for $r_{\ast} = r/h_{\rm e} =0$ as a function of $z_{\ast} = z/h_{\rm e}$, where grey areas correspond to the undermost layers of the eddies}
\label{fig2}
\end{figure}
%____________________________________________________

To justify the distribution of the eddy size $n_{\rm e}(h_{\rm e}) = N_{\rm e} h_{\rm e}^{-3}$, the constantness of $\langle -uw \rangle$ is reproduced in the limit $z/ \delta \rightarrow 0$. That is, for $r=0$ in equation (\ref{eq2_2a}),
%____________________________________________________
\begin{equation}
\label{eq2_3}
\frac{\langle -uw(z) \rangle}{u_{\tau}^2} = -N_{\rm e} \! \int^1_{z/ \delta} \! \frac{dz_{\ast}}{z_{\ast}} \, I_{uw}(0,z_{\ast}) \rightarrow \mbox{const}.
\end{equation}
%____________________________________________________
We have used $z_{\ast} = z/h_{\rm e}$ and $dz_{\ast} /z_{\ast} = -dh_{\rm e} /h_{\rm e}$. The condition of equation (\ref{eq2_1b}) yields $I_{uw}(0,z_{\ast}) = a_{\rm e} z_{\ast}$ with a constant $a_{\rm e}$ at $z_{\ast} \ll 1$. If $z/\delta$ lies in this range of $z_{\ast}$ as shown in figure \ref{fig2}$b$, the integral is equal to $b_{\rm e} - a_{\rm e} z/ \delta$. The constant $b_{\rm e} = \! \int^1_0 (dz_{\ast}/z_{\ast}) \, I_{uw}(0,z_{\ast})$ is dominant in the limit $z/ \delta \rightarrow 0$. To fix the value as $\langle -uw \rangle / u_{\tau}^2 = 1$, we require $N_{\rm e} = -1/b_{\rm e}$, i.e., $N_{\rm e} \varpropto 1/| \mbox{\boldmath{$f$}} |^{2}$ in equation (\ref{eq2_2}) for a given shape of $\mbox{\boldmath{$f$}}$.

The velocity variances $\langle v_i^2(z) \rangle$ such as that in equation (\ref{eq1_1}) are obtained via the same manner. As shown in figures \ref{fig2}$a$ and \ref{fig2}$c$, the wall condition of equation (\ref{eq2_1b}) yields
%____________________________________________________
\begin{equation}
\label{eq2_4}
I_{uu}(0,z_{\ast}) \rightarrow \frac{d_{uu}}{N_{\rm e}}, 
\ \
I_{vv}(0,z_{\ast}) \rightarrow \frac{d_{vv}}{N_{\rm e}}, 
\ \ \mbox{and} \ \ 
I_{ww}(0,z_{\ast}) \varpropto  z_{\ast}^2 
\quad \mbox{as} \ \
z_{\ast} \rightarrow 0.
\end{equation}
%____________________________________________________
Here $d_{uu}$ and $d_{vv}$ are constants. The result for the constant-stress layer $z/\delta \rightarrow 0$ is
%____________________________________________________
\begin{subequations}
\label{eq2_5}
\begin{eqnarray}
\label{eq2_5a}
&& \frac{\langle u^2(z) \rangle}{u_{\tau}^2} = N_{\rm e} \! \int^1_{z/ \delta} \! \frac{dz_{\ast}}{z_{\ast}} \, I_{uu}(0,z_{\ast}) \rightarrow c_{uu}+d_{uu} \ln \! \left( \frac{\delta}{z} \right), \\
\label{eq2_5b}
&&\frac{\langle v^2(z) \rangle}{u_{\tau}^2}  = N_{\rm e} \! \int^1_{z/ \delta} \! \frac{dz_{\ast}}{z_{\ast}} \, I_{vv}(0,z_{\ast}) \rightarrow c_{vv}+d_{vv} \ln \! \left( \frac{\delta}{z} \right), \\
\label{eq2_5c}
&& \frac{\langle w^2(z) \rangle}{u_{\tau}^2} = N_{\rm e} \! \int^1_{z/ \delta} \! \frac{dz_{\ast}}{z_{\ast}} \, I_{ww}(0,z_{\ast}) \rightarrow c_{ww} .
\end{eqnarray}
\end{subequations}
%____________________________________________________
Here $c_{uu}$, $c_{vv}$, and $c_{ww}$ are constants. The law for $\langle u^2(z) \rangle$ has been confirmed with $c_{uu} \simeq$ $1.5$--$2.5$ and $d_{uu} \simeq 1.3$ in a variety of wall turbulence (\cite{hvbs12,marusic13}). We rely on the other laws as well. They do hold in the wall turbulence, albeit not yet certain about the values of $c_{vv}$, $c_{ww}$, and $d_{vv}$ (\cite{sjm13,lm15}).

\section{Velocity correlation along a wall-parallel line} \label{sec3}

Two-point velocity correlations along a wall-parallel line $\langle v_i(x+r,z)v_i(x,z) \rangle$ are studied at a height $z$ in the constant-stress layer. Although the line has been set streamwise, the same discussion is applicable to the case of a spanwise line. We make use of the correlation length
%____________________________________________________
\begin{equation}
\label{eq3_1}
L_{x:v_i v_i}(z) = \frac{\int^{\infty}_0 \! dr \, \langle v_i(x+r,z)v_i(x,z) \rangle}{\langle v_i^2(z) \rangle} .
\end{equation}
%____________________________________________________
This is expected to be well-defined in the attached-eddy hypothesis, where any correlation is determined by the individual eddies that are not dependent on one another and are of finite size. From equation (\ref{eq2_2a}) with $r_{\ast} = r/h_{\rm e}$ and $z_{\ast} = z/h_{\rm e}$,
%____________________________________________________
\begin{subequations}
\label{eq3_2}
\begin{equation}
\label{eq3_2a}
L_{x:v_i v_i}(z) = \frac{\int^{\infty}_0 \! dr \!                     \int^{\delta}_z (dh_{\rm e}/ h_{\rm e})  \, I_{v_i v_i} (   r/h_{\rm e}, z/h_{\rm e})}
                  {                                                   \int^{\delta}_z (dh_{\rm e}/ h_{\rm e})  \, I_{v_i v_i} (       0, z/h_{\rm e})}
               = z \frac{\int^1_{z/\delta} (dz_{\ast} /z_{\ast}^2) \! \int^{\infty}_0 \! dr_{\ast} \, I_{v_i v_i} (r_{\ast}, z_{\ast})}
                        {\int^1_{z/\delta} (dz_{\ast} /z_{\ast}  )                                 \, I_{v_i v_i} (       0, z_{\ast})}.
\end{equation}
%____________________________________________________
The correlation length is also defined in an eddy of a particular size $h_{\rm e}$, 
%____________________________________________________
\begin{equation}
\label{eq3_2b}
{\mit\Lambda}_{x:v_iv_i}(z,h_{\rm e}) = \frac{\int^{\infty}_0 \! dr        \, I_{v_iv_i} (   r/h_{\rm e}, z/h_{\rm e})}
                                             {                                I_{v_iv_i} (       0, z/h_{\rm e})}
                            = h_{\rm e} \frac{\int^{\infty}_0 \! dr_{\ast} \, I_{v_iv_i} (r_{\ast}, z/h_{\rm e})}
                                             {                                I_{v_iv_i} (       0, z/h_{\rm e})}.
\end{equation}
\end{subequations}
%____________________________________________________
While $L_{x:v_i v_i}/z$ is dependent on $z/\delta$ alone, ${\mit\Lambda}_{x:v_iv_i}/h_{\rm e}$ is dependent on $z/h_{\rm e}$ alone.\footnote{
Although equation (\ref{eq3_2a}) could be rewritten so that $L_{x:v_i v_i}/\delta$ is another function of $z/\delta$, it is not considered here because $\delta$ is not a characteristic of the constant-stress layer.}
 That is, $L_{x:v_i v_i}$ and ${\mit\Lambda}_{x:v_iv_i}$ are to be determined respectively by $z/\delta$ and $z/h_{\rm e}$ through $L_{x:v_i v_i}/z$ and ${\mit\Lambda}_{x:v_iv_i}/h_{\rm e}$.

We require $L_{x:v_i v_i}/z$ to be finite. If $L_{x:v_i v_i}/z$ were not finite, it would follow that $L_{x:v_i v_i}$ is not determined and is not well-defined in the attached-eddy hypothesis. The finiteness is also required for ${\mit\Lambda}_{x:v_iv_i}/h_{\rm e}$ because the streamwise size of an eddy is finite as compared with its wall-normal size. For the undermost layer of such an eddy $z_{\ast} = z/h_{\rm e} \ll 1$, there is a constant $\alpha \ge 0$ in relation to equation (\ref{eq3_2b}),
%____________________________________________________
\begin{equation}
\label{eq3_3}
\frac{{\mit\Lambda}_{x:v_iv_i}(z,h_{\rm e})}{h_{\rm e}} \varpropto z_{\ast}^{\alpha}
\! \quad \mbox{and hence} \ \ \! \!
\int^{\infty}_0 \! \! dr_{\ast} \, I_{v_iv_i} (r_{\ast},z_{\ast}) \varpropto z_{\ast}^{\alpha} I_{v_iv_i} (0,z_{\ast})
\! \quad \mbox{as} \ \ 
z_{\ast} \rightarrow 0.
\end{equation}
%____________________________________________________
The asymptotic form of $I_{v_iv_i} (0,z_{\ast})$ is given in equation (\ref{eq2_4}). By substituting the resultant form of $\int^{\infty}_0 \! dr_{\ast} \, I_{v_iv_i} (r_{\ast},z_{\ast})$ into equation (\ref{eq3_2a}), along with $\int^1_{z/\delta} (dz_{\ast} /z_{\ast}) \, I_{v_iv_i} (0,z_{\ast})$ given in equation (\ref{eq2_5}), we obtain $L_{x:v_iv_i}/z$ in the limit $z/\delta \rightarrow 0$. The minimum $\alpha$ value for $L_{x:v_iv_i}/z$ to be finite is adopted as the leading order to determine the asymptotic form of $I_{v_iv_i} (r_{\ast},z_{\ast})$ in equation (\ref{eq3_3}). It determines the asymptotic form of $\langle v_i(x+r,z)v_i(x,z) \rangle$ in the constant-stress layer $z/\delta \rightarrow 0$ via a manner similar to that for the variance $\langle v_i^2(z) \rangle$ in equation (\ref{eq2_5}).

The individual cases of $v_i = u$, $v$, and $w$ are as follows. Only a leading-order behaviour for $z/\delta \rightarrow 0$ is studied about $\int^1_{z/\delta} (dz_{\ast} /z_{\ast}^2) \! \int^{\infty}_0 \! dr_{\ast} \, I_{v_iv_i} (r_{\ast},z_{\ast})$ and $\int^1_{z/\delta} (dz_{\ast} /z_{\ast}) \, I_{v_iv_i} (0,z_{\ast})$, e.g., $\ln (\delta /z)$ about $c_{uu}+d_{uu} \ln (\delta /z)$ in equation (\ref{eq2_5a}).

For the streamwise velocity $u$, we adopt $\alpha = 1$ in equation (\ref{eq3_3}) and use equation (\ref{eq2_4}) to obtain $\int^{\infty}_0 \! dr_{\ast} \, I_{uu} (r_{\ast},z_{\ast}) \varpropto z_{\ast}$ in the limit $z_{\ast} \rightarrow 0$. This is applied for $z/\delta \rightarrow 0$ as
%____________________________________________________
\begin{equation}
\int^1       _{z/\delta} \! \frac{dz_{\ast}}{z_{\ast}^2} \! \int^{\infty}_0 \! \! dr_{\ast} \, I_{uu} (r_{\ast},z_{\ast}) \varpropto \ln \! \left( \frac{\delta}{z} \right) . \nonumber
\end{equation}
%____________________________________________________
On the other hand, from equation (\ref{eq2_5a}),
%____________________________________________________
\begin{equation}
\int^1_{z/\delta}        \! \frac{dz_{\ast}}{z_{\ast}}                                      \, I_{uu} (0,       z_{\ast}) \varpropto \ln \! \left( \frac{\delta}{z} \right) . \nonumber
\end{equation}
%____________________________________________________
By substituting them into equation (\ref{eq3_2a}), we confirm that $L_{x:uu}/z$ is finite. If $\alpha < 1$, it would diverge as $(\delta/z)^{1-\alpha}/ \ln (\delta /z)$. The above asymptotic form of $\int^{\infty}_0 \! dr_{\ast} \, I_{uu} (r_{\ast},z_{\ast}) \varpropto z_{\ast}$ is equivalent through $r_{\ast} = r/h_{\rm e}$ and $z_{\ast}  = z/h_e$ to
%____________________________________________________
\begin{equation}
\int^{\infty}_0 \! \! dr \, I_{uu} \left( \frac{r}{h_{\rm e}} ,\frac{z}{h_{\rm e}} \right) \varpropto z
\! \quad \mbox{as} \ \ 
\frac{z}{h_{\rm e}} \rightarrow 0.
\nonumber
\end{equation}
%____________________________________________________
Thus, $I_{uu} (r/h_{\rm e} ,z/h_{\rm e} )$ in this limit $z/h_{\rm e} \rightarrow 0$ is independent of $h_{\rm e}$ but is only a function of $r/z$, say, $D_{x:uu} (r/z)/N_{\rm e}$. It is independent of $z_{\ast} = z/h_{\rm e}$ if $r/z$ is fixed as $I_{uu} (r/h_{\rm e} ,z/h_{\rm e} ) = I_{uu} (r z_{\ast}/z, z_{\ast})$. The correlation function in the constant-stress layer $z/\delta \rightarrow 0$ is thereby derived from
%____________________________________________________
\begin{equation}
\frac{\langle u(x+r,z)u(x,z) \rangle}{u_{\tau}^2}
=
N_{\rm e} \! \int^1_{z/ \delta} \! \frac{dz_{\ast}}{z_{\ast}} \, I_{uu} \! \left( \frac{r}{z} z_{\ast},z_{\ast} \right). \nonumber
\end{equation}
%____________________________________________________
The result is
%____________________________________________________
\begin{subequations}
\label{eq3_4}
\begin{equation}
\label{eq3_4a}
\frac{\langle u(x+r,z)u(x,z) \rangle}{u_{\tau}^2}
\rightarrow 
C_{x:uu} \! \left( \frac{r}{z} \right) + \ln \! \left( \frac{\delta}{z} \right) D_{x:uu} \! \left( \frac{r}{z} \right). \ 
\end{equation}
%____________________________________________________
Here $C_{x:uu}(r/z)$ is a function of $r/z$. The correlation function of the spanwise velocity $v$ is derived via the same manner,
%____________________________________________________
\begin{equation}
\label{eq3_4b}
\frac{\langle v(x+r,z)v(x,z) \rangle}{u_{\tau}^2}
\rightarrow 
C_{x:vv} \! \left( \frac{r}{z} \right) + \ln \! \left( \frac{\delta}{z} \right) D_{x:vv} \! \left( \frac{r}{z} \right). \
\end{equation}
%____________________________________________________
For the wall-normal velocity $w$, we adopt $\alpha = 0$ in equation (\ref{eq3_3}) and use equation (\ref{eq2_4}) to obtain $\int^{\infty}_0 \! dr_{\ast} \, I_{ww} (r_{\ast},z_{\ast}) \varpropto z_{\ast}^2$ in the limit $z_{\ast} \rightarrow 0$. This is applied for $z/\delta \rightarrow 0$ as
%____________________________________________________
\begin{equation}
\int^1       _{z/\delta} \!    \frac{dz_{\ast}}{z_{\ast}^2} \! \int^{\infty}_0 \! \! dr_{\ast} \, I_{ww} (r_{\ast},z_{\ast}) \rightarrow \mbox{const} . \nonumber
\end{equation}
%____________________________________________________
On the other hand, from equation (\ref{eq2_5c}),
%____________________________________________________
\begin{equation}
\int^1_{z/\delta}        \!    \frac{dz_{\ast}}{z_{\ast}}                                      \, I_{ww} (0,       z_{\ast}) \rightarrow \mbox{const} . \nonumber
\end{equation}
%____________________________________________________
By substituting them into equation (\ref{eq3_2a}), we confirm that $L_{x:ww}/z$ is finite. The above form of $\int^{\infty}_0 \! dr_{\ast} \, I_{ww} (r_{\ast},z_{\ast}) \varpropto z_{\ast}^2$ in the limit $z_{\ast} \rightarrow 0$ leads to $\beta \ge 2$ for $I_{ww} (r_{\ast},z_{\ast}) \varpropto z_{\ast}^{\beta}$ at each of $r_{\ast}$ in this limit. Also, as for $I_{ww} (r_{\ast},z_{\ast}) \varpropto r_{\ast}^{\gamma}$ at each of $z_{\ast}$ in the limit $r_{\ast} \rightarrow 0$, we require $\gamma \ge 0$ because $I_{ww} (0,z_{\ast})$ is finite. With use of $I_{ww} (r z_{\ast}/z, z_{\ast}) \varpropto z_{\ast}^{\beta+\gamma}$ in case of $z_{\ast} \ll 1$,
%____________________________________________________
\begin{equation}
\frac{\langle w(x+r,z)w(x,z) \rangle}{u_{\tau}^2}
=
N_{\rm e} \! \int^1_{z/ \delta} \! \frac{dz_{\ast}}{z_{\ast}} \, I_{ww} \! \left( \frac{r}{z} z_{\ast},z_{\ast} \right), \nonumber
\end{equation}
%____________________________________________________
and hence
%____________________________________________________
\begin{equation}
\label{eq3_4c}
\frac{\langle w(x+r,z)w(x,z) \rangle}{u_{\tau}^2}
\rightarrow 
C_{x:ww} \! \left( \frac{r}{z} \right) .
\end{equation}
\end{subequations}
%____________________________________________________
To relate these correlations continuously to the variances $\langle v_i^2(z) \rangle$ in equation (\ref{eq2_5}), we require $C_{x:v_iv_i}(0) = c_{v_iv_i}$ and $D_{x:v_iv_i}(0) = d_{v_iv_i}$. Especially if $d_{v_iv_i}$ is existent and $\langle v_i^2 (z) \rangle$ is dependent on $\ln (\delta /z)$, then $D_{x:v_iv_i}(r/z)$ is existent and $\langle v_i(x+r,z)v_i(x,z) \rangle$ is dependent on $\ln (\delta /z)$.

The functions $C_{x:v_iv_i}(r/z)$ and $D_{x:v_iv_i}(r/z)$ are to reflect the internal structures of the individual eddies. Since $D_{x:v_iv_i}(r/z)$ is due to eddies of sizes $h_{\rm e}$ up to the height $\delta$ of the turbulence, it is multiplied by $\ln (\delta /z)$ and is existent only for the wall-parallel velocities $u$ and $v$ that are not blocked by the wall. There is also $C_{x:v_iv_i}(r/z)$ due to eddies of sizes $h_{\rm e}$ comparable to the height $z$, existent for all the velocities $u$, $v$, and $w$.

The functional forms of equation (\ref{eq3_4}) hold for distances $r$ down to those in the inertial range. Below this range, there lies the dissipation range where the kinematic viscosity $\nu$ is important. Since we have taken the limit $\nu \rightarrow 0$, those forms do not hold. The velocity fluctuations in the dissipation range are regarded to have been coarse-grained at a length scale in the inertial range (see also \S\ref{sec62}).

Lastly, we reconsider the correlation lengths in the constant-stress layer $z/\delta \rightarrow 0$, by substituting equation (\ref{eq3_4}) into equation (\ref{eq3_1}) with $r_{\star} = r/z$,
%____________________________________________________
\begin{subequations}
\label{eq3_5}
\begin{eqnarray}
\label{eq3_5a}
& & L_{x:uu}(z) \rightarrow z \frac{\int^{\infty}_0 \! dr_{\star} \, C_{x:uu}(r_{\star}) + \ln ( \delta /z) \int^{\infty}_0 \! dr_{\star} \, D_{x:uu}(r_{\star})}
                                   {                                 C_{x:uu}(0)+          \ln ( \delta /z)                               D_{x:uu}(0)} ,        \\
\label{eq3_5b}
& & L_{x:vv}(z) \rightarrow z \frac{\int^{\infty}_0 \! dr_{\star} \, C_{x:vv}(r_{\star}) + \ln ( \delta /z) \int^{\infty}_0 \! dr_{\star} \, D_{x:vv}(r_{\star})}
                                   {                                 C_{x:vv}(0)+          \ln ( \delta /z)                               D_{x:vv}(0)} ,        \\
\label{eq3_5c}
& & L_{x:ww}(z) \rightarrow z \frac{\int^{\infty}_0 \! dr_{\star} \, C_{x:ww}(r_{\star})}
                                   {                                 C_{x:ww}(0)} .
\end{eqnarray}
\end{subequations}
%____________________________________________________
They tend to increase with an increase in the height $z$. However, $L_{x:uu}(z)$ and $L_{x:vv}(z)$ are not proportional to the height $z$ because of the factor $\ln (\delta /z)$. This is not the case for $L_{x:ww}(z)$, which is exactly proportional to the height $z$.

%____________________________________________________
\begin{table}
\begin{center}
\def~{\hphantom{0}}
\begin{tabular}{ccc}
  correlation
& functional form 
& equation 
\\[3pt]
  ${\displaystyle \frac{\langle u(\mbox{\boldmath $x$}+\mbox{\boldmath $r$}_l) u(\mbox{\boldmath $x$}) \rangle}{u_{\tau}^2}    }$ 
& ${\displaystyle C_{l:uu} \! \left( \frac{r}{z} \right) + \ln \! \left( \frac{\delta}{z} \right) D_{l:uu} \! \left( \frac{r}{z} \right)}$ & (\ref{eq3_4a}) and (\ref{eq4_4a}) 
\\[9pt]
  ${\displaystyle \frac{\langle v(\mbox{\boldmath $x$}+\mbox{\boldmath $r$}_l) v(\mbox{\boldmath $x$}) \rangle}{u_{\tau}^2}    }$ 
& ${\displaystyle C_{l:vv} \! \left( \frac{r}{z} \right) + \ln \! \left( \frac{\delta}{z} \right) D_{l:vv} \! \left( \frac{r}{z} \right)}$ & (\ref{eq3_4b}) and (\ref{eq4_4b}) 
\\[9pt]
  ${\displaystyle \frac{\langle w(\mbox{\boldmath $x$}+\mbox{\boldmath $r$}_l) w(\mbox{\boldmath $x$}) \rangle}{u_{\tau}^2}    }$ 
& ${\displaystyle C_{l:ww} \! \left( \frac{r}{z} \right)                                                                                }$ & (\ref{eq3_4c}) and (\ref{eq4_4c}) 
\\
\end{tabular}
\caption{Correlations over the distance $\mbox{\boldmath $r$}_l = \mbox{\boldmath $r$}_x = (r,0,0)$, $\mbox{\boldmath $r$}_y = (0,r,0)$ or $\mbox{\boldmath $r$}_z = (0,0,-r)$ according to the attached-eddy hypothesis}
\label{tab1}
\end{center}
\end{table}
%____________________________________________________

\section{Velocity correlation along a wall-normal line} \label{sec4}

Two-point velocity correlations are also studied over a wall-normal distance $\mbox{\boldmath{$r$}}_l = \mbox{\boldmath{$r$}}_z = (0,0,-r)$ in the range of $0 \le r < z$. According to the attached-eddy hypothesis (see \S\ref{sec2}), the correlation function is given by
%____________________________________________________
\begin{subequations}
\label{eq4_1}
\begin{equation}
\label{eq4_1a}
\frac{\langle v_i(z-r)v_j(z) \rangle}{u_{\tau}^2}
=
N_{\rm e} \! \int^{\delta}_z \! \frac{dh_{\rm e}}{h_{\rm e}} \, J_{v_i v_j} \! \left( \frac{r}{h_{\rm e}}, \frac{z}{h_{\rm e}} \right),
\end{equation}
%____________________________________________________
with the correlation for eddies of the size $h_{\rm e}$,
%____________________________________________________
\begin{equation}
\label{eq4_1b}
J_{v_i v_j} \! \left( \frac{r}{h_{\rm e}}, \frac{z}{h_{\rm e}} \right) 
= 
\int \! \frac{dx_{\rm e}}{h_{\rm e}} \! \int \! \frac{dy_{\rm e}}{h_{\rm e}} \,
f_{v_i} \! \left( \frac{\mbox{\boldmath{$x$}} + \mbox{\boldmath{$r$}}_z - \mbox{\boldmath{$x$}}_{\rm e} }{h_{\rm e}} \right)
f_{v_j} \! \left( \frac{\mbox{\boldmath{$x$}}                           - \mbox{\boldmath{$x$}}_{\rm e} }{h_{\rm e}} \right) .
\end{equation}
\end{subequations}
%____________________________________________________
Here $J_{v_i v_j} (0,z/h_{\rm e})$ is identical to $I_{v_i v_j} (0,z/h_{\rm e})$. The correlation length is defined by using an integration from $r=0$ to $r=z$ as
%____________________________________________________
\begin{equation}
\label{eq4_2}
L_{z:v_iv_i}(z) = \frac{\int^z_0 \! dr \, \langle v_i(z-r)v_i(z) \rangle}{\langle v_i^2(z) \rangle} .
\end{equation}
%____________________________________________________
From equation (\ref{eq4_1a}) with $r_{\ast} = r/h_{\rm e}$ and $z_{\ast} = z/h_{\rm e}$,
%____________________________________________________
\begin{subequations}
\label{eq4_3}
\begin{equation}
\label{eq4_3a}
L_{z:v_i v_i}(z) = \frac{\int^z_0 \! dr \! \int^{\delta}_z (dh_{\rm e}/ h_{\rm e}) \, J_{v_i v_i} ( r/h_{\rm e}, z/h_{\rm e})}
                        {                  \int^{\delta}_z (dh_{\rm e}/ h_{\rm e}) \, J_{v_i v_i} (     0, z/h_{\rm e})}
               = z \frac{\int^1_{z/\delta} (dz_{\ast} /z_{\ast}^2) \! \int^{z_{\ast}}_0 \! dr_{\ast} \, J_{v_i v_i} (r_{\ast},z_{\ast})}
                        {\int^1_{z/\delta} (dz_{\ast} /z_{\ast}  )                                   \, J_{v_i v_i} (       0,z_{\ast})}.
\end{equation}
%____________________________________________________
The correlation length is also defined in an eddy of a particular size $h_{\rm e}$, 
%____________________________________________________
\begin{equation}
\label{eq4_3b}
{\mit\Lambda}_{z:v_iv_i}(z,h_{\rm e}) = \frac{ \int^z_0       \! dr              \, J_{v_i v_i} (    r/h_{\rm e}, z/h_{\rm e})}
                                             {                                      J_{v_i v_i} (        0, z/h_{\rm e})}
                            = h_{\rm e} \frac{ \int^{z/h_{\rm e}}_0 \! dr_{\ast} \, J_{v_i v_i} ( r_{\ast}, z/h_{\rm e})}
                                             {                                      J_{v_i v_i} (        0, z/h_{\rm e})}.
\end{equation} 
\end{subequations}
%____________________________________________________
We require $L_{z:v_i v_i}/z$ and ${\mit\Lambda}_{z:v_iv_i}/h_{\rm e}$ to be finite (see \S\ref{sec3}). Then, since ${\mit\Lambda}_{z:v_iv_i}/h_{\rm e}$ is finite in the limit $z_{\ast} = z/h_{\rm e} \rightarrow 0$, its asymptotic form is identical to that of ${\mit\Lambda}_{x:v_iv_i}/h_{\rm e}$ in equation (\ref{eq3_3}). For the constant-stress layer $z/\delta \rightarrow 0$, the correlation function is derived via the same manner as for those in equation (\ref{eq3_4}),
%____________________________________________________
\begin{equation}
\frac{\langle v_i(z-r)v_i(z) \rangle}{u_{\tau}^2}
=
N_{\rm e} \! \int^1_{z/ \delta} \! \frac{dz_{\ast}}{z_{\ast}} \, J_{v_i v_i} \! \left( \frac{r}{z} z_{\ast},z_{\ast} \right) . \nonumber
\end{equation}
%____________________________________________________
The result is
%____________________________________________________
\begin{subequations}
\label{eq4_4}
\begin{eqnarray}
\label{eq4_4a}
&& \frac{\langle u(z-r)u(z) \rangle}{u_{\tau}^2}
    \rightarrow 
    C_{z:uu} \! \left( \frac{r}{z} \right) + \ln \! \left( \frac{\delta}{z} \right) D_{z:uu} \! \left( \frac{r}{z} \right), \qquad \quad \\
\label{eq4_4b}
&& \frac{\langle v(z-r)v(z) \rangle}{u_{\tau}^2}
    \rightarrow 
    C_{z:vv} \! \left( \frac{r}{z} \right) + \ln \! \left( \frac{\delta}{z} \right) D_{z:vv} \! \left( \frac{r}{z} \right), \\
\label{eq4_4c}
&& \frac{\langle w(z-r)w(z) \rangle}{u_{\tau}^2}
    \rightarrow 
    C_{z:ww} \! \left( \frac{r}{z} \right) .
\end{eqnarray}
\end{subequations}
%____________________________________________________
Here $C_{z:v_iv_i}$ and $D_{z:v_i v_i}$ are functions of $r/z$ with $C_{z:v_i v_i}(0) = C_{x:v_i v_i}(0) = c_{v_i v_i}$ or with $D_{z:v_i v_i}(0) = D_{x:v_i v_i}(0) = d_{v_i v_i}$. The functional forms of these correlations are summarized in table \ref{tab1}.

\section{Energy spectrum along a wall-parallel line} \label{sec5}

From the velocity correlation along a wall-parallel line $\langle v_i(x+r,z)v_i(x,z) \rangle$, the energy spectrum ${\mit\Phi}_{x:v_i v_i}(k,z)$ at a streamwise wavenumber $k$ is obtained as
%____________________________________________________
\begin{equation}
\label{eq5_1}
{\mit\Phi}_{x:v_i v_i}(k,z) = \frac{2}{\pi} \! \int^{\infty}_0 \! \! dr \, \langle v_i(x+r,z)v_i(x,z) \rangle \cos (k r).
\end{equation}
%____________________________________________________
The substitution of equation (\ref{eq3_4}) into equation (\ref{eq5_1}) leads to the energy spectra of the attached-eddy hypothesis in the constant-stress layer $z/\delta \rightarrow 0$,
%____________________________________________________
\begin{subequations}
\label{eq5_2}
\begin{eqnarray}
\label{eq5_2a}
& & \frac{{\mit\Phi}_{x:uu}(k,z)}{z u_{\tau}^2} \rightarrow \tilde{C}_{x:uu}(k z) + \ln \! \left( \frac{\delta}{z} \right) \tilde{D}_{x:uu}(k z), \\
\label{eq5_2b}
& & \frac{{\mit\Phi}_{x:vv}(k,z)}{z u_{\tau}^2} \rightarrow \tilde{C}_{x:vv}(k z) + \ln \! \left( \frac{\delta}{z} \right) \tilde{D}_{x:vv}(k z), \\
\label{eq5_2c}
& & \frac{{\mit\Phi}_{x:ww}(k,z)}{z u_{\tau}^2} \rightarrow \tilde{C}_{x:ww}(k z) .
\end{eqnarray}
\end{subequations}
%____________________________________________________
Here, we have defined $\tilde{C}_{x:v_i v_i}(k z) = (2/\pi) \! \int^{\infty}_0 \! dr_{\star} \, C_{x:v_i v_i}(r_{\star}) \cos (k z r_{\star})$ and $\tilde{D}_{x:v_i v_i}(k z) = (2/\pi) \! \int^{\infty}_0 \! dr_{\star} \, D_{x:v_i v_i}(r_{\star}) \cos (k z r_{\star})$ with $r_{\star} = r/z$. The functional forms are the same even if the wavenumber is spanwise.

There is another model for the energy spectrum (\cite{pa77}). Within a range of wavenumber $k$, some similarity is assumed among eddies of the streamwise size $1/k$ so that ${\mit\Phi}_{x:uu}(k,z)$ depends only on $u_{\tau}$ and $k$, i.e., ${\mit\Phi}_{x:uu}(k,z) \varpropto u_{\tau}^2/k$. If the range of the similarity is from $k = a_{\rm s}/ \delta$ to $k = b_{\rm s}/z$, where $a_{\rm s}$ and $b_{\rm s}$ are constants, it follows that $\langle u^2(z) \rangle = \! \int^{\infty}_0 \! dk \, {\mit\Phi}_{x:uu}(k,z)$ is dependent on $\ln (\delta /z)$ as in the case of equation (\ref{eq2_5a}). The same law of $u_{\tau}^2/k$ is assumed for ${\mit\Phi}_{x:vv}(k,z)$.

Since equation (\ref{eq2_5}) originates in the attached-eddy hypothesis, the energy spectra of the hypothesis have been related to the law of $u_{\tau}^2/k$ (\cite{pc82}). This law is also used to constrain the functional forms of the correlations of the hypothesis (\cite{dnk06}). However, the existence of such a law remains controversial (e.g., \cite{vgs15}). The law of $u_{\tau}^2/k$ is in fact not consistent with the attached-eddy hypothesis. At any height $z \ll \delta$, we find in equation (\ref{eq5_2}) that ${\mit\Phi}_{x:uu}(k,z)$ and ${\mit\Phi}_{x:vv}(k,z)$ are dominated by $\ln ( \delta /z) \tilde{D}_{x:uu}(k z)$ and $\ln ( \delta /z) \tilde{D}_{x:vv}(k z)$. Their spectral shapes are not allowed to yield another factor of $\ln ( \delta /z)$ for the variances $\langle u^2(z) \rangle$ and $\langle v^2(z) \rangle$. Furthermore, the eddies of the streamwise size $1/k$ are not identical to the attached eddies, which have various sizes $h_{\rm e}$ but could contribute to the same streamwise wavenumber $k$ at each of the height $z$. It is coincidental for equation (\ref{eq2_5}) to be derived from these two.

The similarity law of $u_{\tau}^2/k$ could yet approximate the internal structures of the attached eddies, i.e., $\tilde{C}_{x:v_i v_i}(k z)$ or $\tilde{D}_{x:v_i v_i}(k z)$ in equation (\ref{eq5_2}), if the range of the similarity is not from $k = a_{\rm s} / \delta$ but is from $k = a_{\rm s} /z$. At the height $z \ll \delta$, the premultiplied spectrum $k {\mit\Phi}_{x:uu}(k, z)$ is known to exhibit an energy-containing broad plateau (e.g., \cite{vgs15}). This could be approximated by the law of $u_{\tau}^2/k$ through $\tilde{D}_{x:uu}(k z) \varpropto 1/k z$. From equation (\ref{eq5_1}),
%____________________________________________________
\begin{equation}
\label{eq5_3}
\langle v_i(x+r,z)v_i(x,z) \rangle = \! \int^{\infty}_0 \! \! dk \, {\mit\Phi}_{x:v_i v_i}(k ,z) \cos (k r) .
\end{equation}
%____________________________________________________
There is an asymptotic relation about the cosine integral function,
%____________________________________________________
\begin{subequations}
\begin{equation}
\label{eq5_4a}
\int^{\infty}_T \! \! dt \, \frac{\cos (t)}{t} \rightarrow -\ln |T| - \gamma_{\rm E} \quad \mbox{with Euler's constant $\gamma_{\rm E} = 0.57721 ...$} \quad \mbox{as} \ \ T \rightarrow 0.
\end{equation}
%____________________________________________________
If the distance $r$ is comparable to the height $z$, our law of $u_{\tau}^2/k$ yields a factor of $\ln (r/z)$ as
%____________________________________________________
\begin{equation}
\label{eq5_4b}
\int^{b_{\rm s}/z}_{a_{\rm s}/z} \! \! dk \, \frac{\cos (k r)}{k} \rightarrow -\ln \! \left( \frac{a_{\rm s} r}{z} \right) - \gamma_{\rm E}
\quad \mbox{as} \ \ a_{\rm s} \rightarrow 0 \ \ \mbox{and} \ \ b_{\rm s} \rightarrow \infty.
\end{equation}
\end{subequations}
%____________________________________________________
Thus, the correlation function $\langle u(x+r,z)u(x,z) \rangle$ could exhibit approximate dependence on $\ln (r/z)$ in a range of $r/z$, which is actually seen in the wall turbulence (\cite{dnk06,cmmvs15}, see also \cite{dk14}).

\section{Other correlations} \label{sec6}

Having studied the velocity correlations in table \ref{tab1} by using the minimum assumptions of the attached-eddy hypothesis, some other correlations are studied by adding assumptions that are consistent with the hypothesis. Their functional forms are summarized in table \ref{tab2}.

%____________________________________________________
\begin{table}
\begin{center}
\def~{\hphantom{0}}
\begin{tabular}{ccc}
  correlation
& functional form
& equation 
\\[3pt]
  ${\displaystyle
     \frac{\langle v_i^2(\mbox{\boldmath $x$}+\mbox{\boldmath $r$}_l)                 v_i^2(\mbox{\boldmath $x$}) \rangle
         - \langle v_i^2(\mbox{\boldmath $x$}+\mbox{\boldmath $r$}_l) \rangle \langle v_i^2(\mbox{\boldmath $x$}) \rangle } {u_{\tau}^4}}$
& ${\displaystyle
    \frac{2\langle v_i  (\mbox{\boldmath $x$}+\mbox{\boldmath $r$}_l)                 v_i  (\mbox{\boldmath $x$}) \rangle^2}{u_{\tau}^4}}$ 
& (\ref{eq6_2})
\\[9pt]
  ${\displaystyle 
    \frac{\langle uw (\mbox{\boldmath $x$}+\mbox{\boldmath $r$}_l) uw (\mbox{\boldmath $x$}) \rangle - \langle -uw \rangle^2 }{u_{\tau}^4}}$
& ${\displaystyle C_{l:uw uw} \! \left( \frac{r}{z} \right) + \ln \! \left( \frac{\delta}{z} \right) D_{l:uw uw} \! \left( \frac{r}{z} \right) }$ 
& (\ref{eq6_3})
\\[9pt]
  ${\displaystyle 
    \frac{\langle \varepsilon (\mbox{\boldmath $x$}+\mbox{\boldmath $r$}_l)                 \varepsilon (\mbox{\boldmath $x$}) \rangle
        - \langle \varepsilon (\mbox{\boldmath $x$}+\mbox{\boldmath $r$}_l) \rangle \langle \varepsilon (\mbox{\boldmath $x$}) \rangle }{u_{\tau}^6/z^2}}$
& ${\displaystyle C_{l:\varepsilon \varepsilon} \! \left( \frac{r}{z} \right)}$ 
& (\ref{eq6_8})
\\
\end{tabular}
\caption{Correlations over the distance $\mbox{\boldmath $r$}_l = \mbox{\boldmath $r$}_x = (r,0,0)$, $\mbox{\boldmath $r$}_y = (0,r,0)$ or $\mbox{\boldmath $r$}_z = (0,0,-r)$ according to the attached-eddy hypothesis with some additional assumption}
\label{tab2}
\end{center}
\end{table}
%____________________________________________________

\subsection{Kinetic energy and momentum transfer rate} \label{sec61}

To obtain a moment $\langle v_{i}^{m}(\mbox{\boldmath $x$}+\mbox{\boldmath $r$}_l) v_{j}^{m}(\mbox{\boldmath $x$}) \rangle$ of some order $m$, the corresponding cumulants for eddies of each size $h_{\rm e}$ are integrated from $h_{\rm e} = z$ to $h_{\rm e} = \delta$. Any cumulant of a sum of random variables is equal to the sum of cumulants of the variables if they are independent of one another (e.g., \cite{my71}). The moment of the sum is not equal to the sum of the moments, except for low-order central moments such as $I_{v_i v_j}$ and $J_{v_i v_j}$, defined for $\langle v_i(\mbox{\boldmath $x$}+\mbox{\boldmath $r$}_l) v_j(\mbox{\boldmath $x$}) \rangle$ in equations (\ref{eq2_2}) and (\ref{eq4_1}), which serve also as cumulants. Although a combination of those integrals could yield $\langle v_{i}^{m}(\mbox{\boldmath $x$}+\mbox{\boldmath $r$}_l) v_{j}^{m}(\mbox{\boldmath $x$}) \rangle$, it tends to be a complicated combination and tends to depend on an uncertain value of $N_{\rm e}$. There is a more practical approach.

We set many eddies at each position on the wall, by increasing the number of the eddies $N_{\rm e}$. According to the central limit theorem (e.g., \cite{my71}), the resultant velocity field is approximated to be Gaussian and to be determined by $\langle v_i(\mbox{\boldmath $x$}+\mbox{\boldmath $r$}_l) v_j(\mbox{\boldmath $x$}) \rangle$ alone. The other cumulants are all negligible. This is known as a good approximation for the constant-stress layer of the wall turbulence (\cite{ff96,mm13}).\footnote{
The boundary layer actually exhibits $\langle u^4 \rangle / \langle u^2 \rangle^2 \simeq 2.8$ and $\langle v^4 \rangle / \langle v^2 \rangle^2 \simeq \langle w^4 \rangle / \langle w^2 \rangle^2 \simeq 3.4$, for which the Gaussian value is $3$ (\cite{ff96}).}

The odd-order moments are all equal to $0$ in such a Gaussian field, while any even-order moment is equal to a sum of products of the above correlations. In particular,
%____________________________________________________
\begin{eqnarray}
\label{eq6_1}
   \langle v_{i_1}(\mbox{\boldmath $x$}_1) v_{i_2}(\mbox{\boldmath $x$}_2)                 v_{i_3}(\mbox{\boldmath $x$}_3) v_{i_4}(\mbox{\boldmath $x$}_4) \rangle 
&=&
   \langle v_{i_1}(\mbox{\boldmath $x$}_1) v_{i_2}(\mbox{\boldmath $x$}_2) \rangle \langle v_{i_3}(\mbox{\boldmath $x$}_3) v_{i_4}(\mbox{\boldmath $x$}_4) \rangle  \nonumber \\
&+&
   \langle v_{i_1}(\mbox{\boldmath $x$}_1) v_{i_3}(\mbox{\boldmath $x$}_3) \rangle \langle v_{i_2}(\mbox{\boldmath $x$}_2) v_{i_4}(\mbox{\boldmath $x$}_4) \rangle  \nonumber \\
&+&
   \langle v_{i_1}(\mbox{\boldmath $x$}_1) v_{i_4}(\mbox{\boldmath $x$}_4) \rangle \langle v_{i_2}(\mbox{\boldmath $x$}_2) v_{i_3}(\mbox{\boldmath $x$}_3) \rangle .
\end{eqnarray}
%____________________________________________________
The correlation of the kinetic energy per unit mass $v_i^2 = u^2$, $v^2$, or $w^2$ in the direction $l = x$, $y$, or $z$ is thereby related to that of the velocity $v_i$ in equation (\ref{eq3_4}) or (\ref{eq4_4}),
%____________________________________________________
\begin{equation}
\label{eq6_2}
  \langle v_i^2(\mbox{\boldmath $x$}+\mbox{\boldmath $r$}_l)                 v_i^2(\mbox{\boldmath $x$}) \rangle
- \langle v_i^2(\mbox{\boldmath $x$}+\mbox{\boldmath $r$}_l) \rangle \langle v_i^2(\mbox{\boldmath $x$}) \rangle
=
2 \langle v_i  (\mbox{\boldmath $x$}+\mbox{\boldmath $r$}_l)                 v_i  (\mbox{\boldmath $x$}) \rangle^2 . \\
\end{equation}
%____________________________________________________
From equation (\ref{eq6_1}), we also obtain 
%____________________________________________________
\begin{eqnarray}
   \langle uw(\mbox{\boldmath $x$}+\mbox{\boldmath $r$}_l) uw(\mbox{\boldmath $x$}) \rangle
 - \langle uw(\mbox{\boldmath $x$}+\mbox{\boldmath $r$}_l) \rangle \langle uw(\mbox{\boldmath $x$}) \rangle 
&=&\langle u(\mbox{\boldmath $x$}+\mbox{\boldmath $r$}_l) u(\mbox{\boldmath $x$}) \rangle \langle w(\mbox{\boldmath $x$}+\mbox{\boldmath $r$}_l) w(\mbox{\boldmath $x$}) \rangle \nonumber \\ 
&+&\langle u(\mbox{\boldmath $x$}+\mbox{\boldmath $r$}_l) w(\mbox{\boldmath $x$}) \rangle \langle w(\mbox{\boldmath $x$}+\mbox{\boldmath $r$}_l) u(\mbox{\boldmath $x$}) \rangle \nonumber .
\end{eqnarray}
%____________________________________________________
Since the wall condition of equation (\ref{eq2_1b}) yields $I_{uw}(0,z_{\ast}) \varpropto z_{\ast}$ for $z_{\ast} \rightarrow 0$ (see \S\ref{sec2}), the correlations $\langle u(\mbox{\boldmath $x$}+\mbox{\boldmath $r$}_l) w(\mbox{\boldmath $x$}) \rangle$ and $\langle w(\mbox{\boldmath $x$}+\mbox{\boldmath $r$}_l) u(\mbox{\boldmath $x$}) \rangle$ depend only on $r/z$ in the limit $z/\delta \rightarrow 0$ (see \S\ref{sec3}). By also using equation (\ref{eq3_4}) or (\ref{eq4_4}) for $\langle u(\mbox{\boldmath $x$}+\mbox{\boldmath $r$}_l) u(\mbox{\boldmath $x$}) \rangle$ and $\langle w(\mbox{\boldmath $x$}+\mbox{\boldmath $r$}_l) w(\mbox{\boldmath $x$}) \rangle$, the correlation function of the rate of momentum transfer per unit mass $-uw$ is derived for the constant-stress layer $z/ \delta \rightarrow 0$ as
%____________________________________________________
\begin{subequations}
\label{eq6_3}
\begin{eqnarray}
&&  \frac{ \langle uw(x+r,z) uw(x,z) \rangle - \langle -uw \rangle^2 } {u_{\tau}^4} 
    \rightarrow 
    C_{x:uw uw} \! \left( \frac{r}{z} \right) + \ln \! \left( \frac{\delta}{z} \right) D_{x:uw uw} \! \left( \frac{r}{z} \right) , \label{eq6_3a} \quad \\
&&  \frac{ \langle uw(z-r)   uw(z)   \rangle - \langle -uw \rangle^2 } {u_{\tau}^4} 
    \rightarrow 
    C_{z:uw uw} \! \left( \frac{r}{z} \right) + \ln \! \left( \frac{\delta}{z} \right) D_{z:uw uw} \! \left( \frac{r}{z} \right) . \label{eq6_3b}
\end{eqnarray}
\end{subequations}
%____________________________________________________
As for the functions $C_{l:uw uw}(r/z)$ and $D_{l:uw uw}(r/z)$, we require $C_{x:uw uw}(0) = C_{z:uw uw}(0)$ and $D_{x:uw uw}(0) = D_{z:uw uw}(0)$. The fluctuations of $-uw$ are dependent on those of $u$, which are in turn dependent on the factor  $\ln ( \delta /z)$. This is also the case for the correlation along a spanwise line.

\subsection{Energy dissipation rate} \label{sec62}

Since the attached-eddy hypothesis is a model for the energy-containing eddies, it has not been applied to the rate of energy dissipation per unit mass $\varepsilon (\mbox{\boldmath $x$})$. There is nevertheless a possibility for this application, if we add an assumption and if we ignore fluctuations at the smallest length scales.

The dissipation fields of the individual eddies $\varepsilon_{\rm e} (\mbox{\boldmath $x$})$ are assumed to have some identical shape with a common characteristic rate $u_{\tau}^3/h_{\rm e}$,
%____________________________________________________
\begin{equation}
\label{eq6_4}
\varepsilon_{\rm e} (\mbox{\boldmath $x$})
=
\frac{u_{\tau}^3}{h_{\rm e}} f_{\varepsilon} \! \left( \frac{\mbox{\boldmath{$x$}} - \mbox{\boldmath{$x$}}_{\rm e} }{h_{\rm e}} \right) .
\end{equation}
%____________________________________________________
We have the corresponding Kolmogorov length $\eta_{\rm e} = (\nu^3 / \varepsilon_{\rm e} )^{1/4} \varpropto (\nu /u_{\tau})^{3/4} h_{\rm e}^{1/4}$, which is not proportional to the eddy size $h_{\rm e}$. Thus, equation (\ref{eq6_4}) does not hold for fluctuations in the dissipation range around that scale $\eta_{\rm e}$. Such fluctuations are regarded to have been coarse-grained at a length scale in the inertial range.

We set the condition at $z=0$ to be $f_{\varepsilon} \ne  0$, although the last result is the same even in case of $f_{\varepsilon} = 0$. The dissipation rate $\varepsilon (\mbox{\boldmath $x$})$ is averaged at a height $z$ as
%____________________________________________________
\begin{equation}
\langle \varepsilon (\mbox{\boldmath $x$}) \rangle
=
\! \int^{\delta}_z \! \frac{dh_{\rm e}}{h_{\rm e}}
   \left[
          h_{\rm e}^3 \, n_{\rm e}(h_{\rm e}) \!
          \int \! \frac{dx_{\rm e}}{h_{\rm e}} \! 
          \int \! \frac{dy_{\rm e}}{h_{\rm e}} \, \frac{u_{\tau}^3}{h_{\rm e}} f_{\varepsilon} \! \left( \frac{\mbox{\boldmath{$x$}} - \mbox{\boldmath{$x$}}_{\rm e} }{h_{\rm e}} \right)
\right]. 
\nonumber
\end{equation}
%____________________________________________________
By using $n_{\rm e}(h_{\rm e}) = N_{\rm e} h_{\rm e}^{-3}$ with $N_{\rm e} \varpropto 1/f_{\varepsilon}$ for a given shape of $f_{\varepsilon}$ (see \S\ref{sec2}) and also by using $\langle \varepsilon (z) \rangle$ in place of $\langle \varepsilon (\mbox{\boldmath $x$}) \rangle$,
%____________________________________________________
\begin{subequations}
\label{eq6_5}
\begin{equation}
\label{eq6_5a}
\frac{\langle \varepsilon (z) \rangle}{u_{\tau}^3/z}
=
N_{\rm e} \! \int^{\delta}_z \! \frac{dh_{\rm e}}{h_{\rm e}} \, I_{\varepsilon} \! \left( \frac{z}{h_{\rm e}} \right),
\end{equation}
%____________________________________________________
with the average for eddies of the size $h_{\rm e}$,
%____________________________________________________
\begin{equation}
\label{eq6_5b}
I_{\varepsilon} \! \left( \frac{z}{h_{\rm e}} \right) 
= 
\frac{z}{h_{\rm e}} \int \! \frac{dx_{\rm e}}{h_{\rm e}} \! \int \! \frac{dy_{\rm e}}{h_{\rm e}} \,
f_{\varepsilon} \! \left( \frac{\mbox{\boldmath{$x$}} - \mbox{\boldmath{$x$}}_{\rm e} }{h_{\rm e}} \right).
\end{equation}
\end{subequations}
%____________________________________________________
For $z_{\ast} = z/h_{\rm e} \rightarrow 0$, we have $I_{\varepsilon}(z_{\ast}) \varpropto z_{\ast}$ (see figure \ref{fig2}). The mean rate of energy dissipation in the constant-stress layer $z/\delta \rightarrow 0$ is 
%____________________________________________________
\begin{equation}
\label{eq6_6}
\frac{\langle \varepsilon (z) \rangle}{u_{\tau}^3/z}
=
N_{\rm e} \! \int^1_{z/ \delta} \! \frac{dz_{\ast}}{z_{\ast}} \, I_{\varepsilon} ( z_{\ast} ) \rightarrow \mbox{const}.
\end{equation}
%____________________________________________________
To determine the coefficient for $f_{\varepsilon} \varpropto 1/N_{\rm e}$, the above constant $N_{{\rm e} \!} \int^1_0 (dz_{\ast} /z_{\ast}) \, I_{\varepsilon} ( z_{\ast} )$ is set equal to the inverse of the von K\'arm\'an constant, $1/ \kappa \simeq 2.5$. It corresponds to a result of the local equilibrium approximation (\cite{t61}), where the constant-stress layer is under an equilibrium between local rates of production and dissipation of the turbulence kinetic energy.\footnote{
For the actual constant-stress layer, it is considered that the local rate of energy dissipation is $90$--$100$\% of the local rate of energy production (\cite{lm15}).}

The correlation function over a streamwise distance $\mbox{\boldmath $r$}_l = \mbox{\boldmath $r$}_x = (r,0,0)$ at a height $z$ is given by
%____________________________________________________
\begin{subequations}
\label{eq6_7}
\begin{equation}
\label{eq6_7a}
\frac{\langle \varepsilon (x+r,z) \varepsilon (x,z) \rangle - \langle \varepsilon (z) \rangle^2}{u_{\tau}^6/z^2}
=
N_{\rm e} \! \int^{\delta}_z \! \frac{dh_{\rm e}}{h_{\rm e}} \left[ I_{\varepsilon \varepsilon}   \! \left( \frac{r}{h_{\rm e}}, \frac{z}{h_{\rm e}} \right) 
                                                                  - I_{\varepsilon}            ^2 \! \left(                \frac{z}{h_{\rm e}} \right) \right],
\end{equation}
%____________________________________________________
with the correlation $I_{\varepsilon \varepsilon}-I_{\varepsilon}^2$ for eddies of the size $h_{\rm e}$, which is made up from $I_{\varepsilon}(z/h_{\rm e})$ in equation (\ref{eq6_5b}) and also from
%____________________________________________________
\begin{equation}
\label{eq6_7b}
I_{\varepsilon \varepsilon} \! \left( \frac{r}{h_{\rm e}}, \frac{z}{h_{\rm e}} \right) 
= 
\frac{z^2}{h_{\rm e}^2} \int \! \frac{dx_{\rm e}}{h_{\rm e}} \! \int \! \frac{dy_{\rm e}}{h_{\rm e}} \,
f_{\varepsilon} \! \left( \frac{\mbox{\boldmath{$x$}} + \mbox{\boldmath{$r$}}_x - \mbox{\boldmath{$x$}}_{\rm e} }{h_{\rm e}} \right)
f_{\varepsilon} \! \left( \frac{\mbox{\boldmath{$x$}}                           - \mbox{\boldmath{$x$}}_{\rm e} }{h_{\rm e}} \right).
\end{equation}
\end{subequations}
%____________________________________________________
For $z_{\ast} = z/h_{\rm e} \rightarrow 0$, we have $I_{\varepsilon \varepsilon} (0,z_{\ast}) - I_{\varepsilon}^2 (z_{\ast}) \varpropto z_{\ast}^2$ (see figure \ref{fig2}). Via a manner similar to that for the velocity correlations (see \S\ref{sec3}),
%____________________________________________________
\begin{equation}
\frac{\langle \varepsilon (x+r,z) \varepsilon (x,z) \rangle - \langle \varepsilon (z) \rangle^2}{u_{\tau}^6/z^2}
=
N_{\rm e} \! \int^1_{z/ \delta} \! \frac{dz_{\ast}}{z_{\ast}} \left[  I_{\varepsilon \varepsilon}   \! \left( \frac{r}{z} z_{\ast},z_{\ast} \right) 
                                                                     -I_{\varepsilon}            ^2 \! \left(                      z_{\ast} \right) \right] , \nonumber
\end{equation}
%____________________________________________________
and hence
%____________________________________________________
\begin{subequations}
\label{eq6_8}
\begin{equation}
\label{eq6_8a}
\frac{\langle \varepsilon (x+r,z) \varepsilon (x,z) \rangle - \langle \varepsilon (z) \rangle^2}{u_{\tau}^6/z^2}
\rightarrow 
C_{x:\varepsilon \varepsilon} \! \left( \frac{r}{z} \right) .
\end{equation}
%____________________________________________________
The same form is derived for the correlation function over a spanwise distance. For that over a wall-normal distance $\mbox{\boldmath $r$}_l = \mbox{\boldmath $r$}_z = (0,0,-r)$, we replace $\mbox{\boldmath{$r$}}_x$ with $\mbox{\boldmath{$r$}}_z$ in equation (\ref{eq6_7b}) to define $J_{\varepsilon \varepsilon}(r/h_{\rm e},z/h_{\rm e})$. Then,
%____________________________________________________
\begin{equation}
\frac{\langle \varepsilon (z-r) \varepsilon (z) \rangle - \langle \varepsilon (z-r) \rangle \langle \varepsilon (z) \rangle}{u_{\tau}^6/z^2}
=
N_{\rm e} \! \int^1_{z/ \delta} \! \frac{dz_{\ast}}{z_{\ast}} \left[  J_{\varepsilon \varepsilon}  \! \left( \frac{r}{z} z_{\ast}, z_{\ast} \right) 
                                                                     -I_{\varepsilon}              \! \left( z_{\ast}- \frac{r}{z} z_{\ast} \right) 
                                                                      I_{\varepsilon}              \! \left(                       z_{\ast} \right) \right]. \nonumber
\end{equation}
%____________________________________________________
The result is
%____________________________________________________
\begin{equation}
\label{eq6_8b}
\frac{\langle \varepsilon (z-r) \varepsilon (z) \rangle - \langle \varepsilon (z-r) \rangle \langle \varepsilon (z) \rangle}{u_{\tau}^6/z^2}
\rightarrow 
C_{z:\varepsilon \varepsilon} \! \left( \frac{r}{z} \right) .
\end{equation}
\end{subequations}
%____________________________________________________
Here $C_{l:\varepsilon \varepsilon}$ is a function of $r/z$ with $C_{x:\varepsilon \varepsilon}(0) = C_{y:\varepsilon \varepsilon}(0) = C_{z:\varepsilon \varepsilon}(0)$. Since the rate $\varepsilon_{\rm e}$ of an eddy is inversely proportional to its size $h_{\rm e}$, the entire rate $\varepsilon$ is independent of $\ln ( \delta /z)$ that originates in eddies of sizes $h_{\rm e}$ much larger than the height $z$. On the other hand, from $\varepsilon_{\rm e} \varpropto f_{\varepsilon} \varpropto 1/N_{\rm e}$, we have $C_{l:\varepsilon \varepsilon}(r/z) \varpropto 1/N_{\rm e}$. If $N_{\rm e}$ were too large, the fluctuations of the entire rate $\varepsilon$ would be negligible. They are actually not negligible even over large distances $r$ in the wall turbulence (e.g., \cite{mouri06}).

\section{Concluding discussion} \label{sec7}

Two-point velocity correlations have been studied theoretically for the constant-stress layer of wall turbulence on the basis of the attached-eddy hypothesis, i.e., a phenomenological model of a random superposition of energy-containing eddies that are attached to the wall (\cite{t76}). While the shapes of the eddies are set identical to one another with a common characteristic velocity $u_{\tau}$, their sizes are distributed up to a finite value in each of the directions.

From the random distribution and the finite sizes of the attached eddies, it follows that the correlation lengths defined in equations (\ref{eq3_2}) and (\ref{eq4_3}) are always existent. We have used this fact to derive the most general forms of the correlation functions in equations (\ref{eq3_4}) and (\ref{eq4_4}) and of the energy spectra in equation (\ref{eq5_2}). The results are summarized in table \ref{tab1}.

The correlation functions $\langle v_i(\mbox{\boldmath $x$}+\mbox{\boldmath $r$}_l) v_i(\mbox{\boldmath $x$}) \rangle$ are made from $C_{l:v_i v_i}(r/z)$ due to eddies of sizes $h_{\rm e}$ comparable to the height $z$ and from $D_{l:v_i v_i}(r/z)$ due to eddies of sizes $h_{\rm e}$ larger than the height $z$. We have obtained $D_{l:v_i v_i}(r/z)$ only for the wall-parallel velocities $u$ and $v$ that are not blocked by the wall. It is multiplied by $\ln (\delta /z)$ because the size $h_{\rm e}$ of the large eddies is up to the height $\delta$ of the turbulence.

These results are not consistent with the results of the previous studies, e.g., the spectra in the form of ${\mit\Phi}_{x:v_i v_i}(k, z) \varpropto u_{\tau}^2/k$ (\cite{pc82}). While we have focused on the minimum assumptions of the attached-eddy hypothesis, the previous studies had invoked additional assumptions. They are in fact not consistent with the hypothesis, aside from which is more reliable as a model for the wall turbulence.

Without any confounding effect of such additional assumptions, our results would offer an opportunity to assess the extent to which the attached-eddy hypothesis is consistent with the actual wall turbulence. However, at present, the direct numerical simulations are not at Reynolds numbers that are high enough to attain a layer of the strictly constant stress (\cite{sjm13,lm15}). The experiments are yet based mostly on Taylor's frozen-eddy hypothesis. Improvements are desirable for the simulations and for the experiments.

There are nevertheless some experimental implications. It is known that the value of $d_{uu} = D_{l:uu}(0)$ is common to a variety of flows (\cite{marusic13}). This commonness is expected for the entire function $D_{l:uu}(r/z)$, which reflects the undermost layers of the attached eddies that are insensitive to the flow configuration. The value of $c_{uu} = C_{l:uu}(0)$ is not common, and so is not the entire function $C_{l:uu}(r/z)$. We expect these implications to apply further to the other velocities $v$ and $w$.

The two-point correlation of the wall-normal velocity $w$ is determined by the attached eddies of sizes $h_{\rm e}$ comparable to the height $z$ of the two points. If such eddies are actually dominant in the wall turbulence (\cite{a07}), their internal structures would be clarified in that correlation, especially about whether the shape of the eddy is independent of its size. This is not the case for the correlations of the wall-parallel velocities $u$ and $v$, which are affected by the larger eddies.

By adding assumptions that are consistent with the attached-eddy hypothesis, we have derived the correlations of some other quantities. That is, if the velocity field is Gaussian, the correlation functions of the kinetic energies $u^2$, $v^2$, and $w^2$ and that of the momentum transfer rate $-uw$ are given in equations (\ref{eq6_2}) and (\ref{eq6_3}). If the energy dissipation rate of an attached eddy $\varepsilon_{\rm e}$ is proportional to $u_{\tau}^3/h_{\rm e}$, the correlation function of the entire rate $\varepsilon$ is given in equation (\ref{eq6_8}). The results are summarized in table \ref{tab2}.

The logarithmic law such as that in equation (\ref{eq1_1}) has been found also for the variance of the pressure fluctuations $p$ (\cite{jm08,sjm13}). If this law is due to the attached eddies, the two-point correlation $\langle p(\mbox{\boldmath $x$}+\mbox{\boldmath $r$}_l) p(\mbox{\boldmath $x$}) \rangle$ is expected to have a functional form that is identical to those of the wall-parallel velocities $u$ and $v$ given in equations (\ref{eq3_4}) and (\ref{eq4_4}).

Lastly, we note that these results would allow us to use the attached-eddy hypothesis in its most general form. An example is the coarse-graining to characterize or model the wall turbulence, say, the atmospheric boundary layer in a particular class of numerical simulations (\cite{w04}). Over a streamwise length $R$, we coarse-grain a quantity $q$ as 
%____________________________________________________
\begin{equation}
\label{eq7_1}
q_R(x,y,z) = \frac{1}{R} \! \int^{R}_{0}  \! \! dr \, q(x+r,y,z).
\end{equation}
%____________________________________________________
Through a relation (e.g., \cite{my71}), the variance of $q_R$ is dependent on the two-point correlation of $q$,
%____________________________________________________
\begin{equation}
\langle q^2_R(z) \rangle - \langle q_R (z) \rangle^2
= 
\frac{2}{R^2} \! \int^{R}_{0} \! \! dr_1 \! \int^{r_1}_{0} \! \! dr_2 \, \left[ \langle q(x+r_2,z) q(x,z) \rangle - \langle q(z) \rangle^2 \right], \nonumber
\end{equation}
%____________________________________________________
and hence
%____________________________________________________
\begin{subequations}
\begin{equation}
\label{eq7_1}
\langle q^2_R(z) \rangle - \langle q_R (z) \rangle^2
= 
\frac{2}{R^2} \! \int^{R}_{0} \! \! dr \, (R-r) \left[ \langle q(x+r,z) q(x,z) \rangle - \langle q(z) \rangle^2 \right].
\end{equation}
%____________________________________________________
By using equation (\ref{eq3_1}) to define the correlation length $L_{x:qq}(z)$,
%____________________________________________________
\begin{equation}
\label{eq7_1}
\langle q^2_R(z) \rangle - \langle q_R (z) \rangle^2
\rightarrow 
\frac{2 L_{x:qq}(z)}{R} \left[ \langle q^2(z) \rangle - \langle q(z) \rangle^2 \right]
\quad \mbox{as} \ \
R \rightarrow \infty.
\end{equation}
\end{subequations}
%____________________________________________________
If the two-point correlation is a function of $r/z$ alone as for $q = w$, $w^2$, and $\varepsilon$ in tables \ref{tab1} and \ref{tab2}, the variance is a function of $R/z$ alone. The remaining variances depend also on $\ln (\delta/z)$. Such a constraint would in turn constrain the theory or model on that coarse-graining. Thus, for this and other various studies of wall turbulence, useful would be the general forms of the correlation functions derived here from the minimum assumptions of the attached-eddy hypothesis.

\section*{Acknowledgments}
This work was supported in part by KAKENHI Grant No. 25340018. The author is grateful to anonymous referees for stimulating comments.

\end{document}